# All-organic self-separating three-dimensionally nanoarchitected electrochemical energy storage devices

William R. T. Tait[1,2], Sriram Murali[1], Chao-Hua Hsu[1], Jantakan Nedsaengtip[2], Christina Lee[1], R. Paxton Thedford[2], Vibha Kalra[2], Joerg G. Werner[3,4], Ulrich B. Wiesner[1,5,6,]*

[1] Department of Materials Science and Engineering, Cornell University, Ithaca, New York 14853, United States
[2] Robert Frederick Smith School of Chemical and Biomolecular Engineering, Cornell University, Ithaca, New York 14853, United States
[3] Department of Mechanical Engineering, Boston University, Boston, Massachusetts 02215, United States
[4] Division of Materials Science and Engineering, Boston University, Boston, Massachusetts 02215, United States
[5] Department of Design Tech, Cornell University; Ithaca, New York 14853, United States
[6] Kavli Institute at Cornell for Nanoscale Science, Cornell University, Ithaca, New York 14853, United States
* Corresponding author

**Abstract**

This work realizes a three-dimensionally (3D) nanoarchitected, all organic, "self-separating" lithium-ion electrochemical energy storage (EES) device that is cycled as a solid-state full cell. The device is enabled by a monolithic carbon anode with a co-continuous pore network, derived from the structure direction of resols by an ultra-large molar mass block copolymer (BCP), poly(styrene-*block*-2-dimethylaminoethyl methacrylate) (SA). Electropolymerization of a single-phase conductive and redox-active material, poly((2,3-dihydrothieno[3,4-b][1,4]dioxin-2-yl)methyl 9,10-dioxo-9,10-dihydroanthracene-2-carboxylate) (PAQEDOT), into the pore space provides the cathode of the cell. The device is electronically contacted to the relevant electrode network enabled by the co-continuous nature of each electrode. Electrochemical processing via cycling against external lithium in an electrolyte generates a solid electrolyte interphase (SEI) as a separator and lithiates the cell electrodes, after which the EES device is cycled in the solid state. While the full cell does not demonstrate high cyclability, the best full cell demonstrates a discharge capacity of 267 milliamp hours per gram (mAh/g). This work marks, to the best of knowledge of the authors, the first example of an all-organic materials derived 3D nanoarchitected EES device, as well as the first design of "self-separating" cell fabrication. Furthermore, generalization of the design to another co-continuous carbon form factor is demonstrated.

**Introduction**

Energy storage is a key facet of the complex solutions to the world's energy demands.[1] Electrochemical energy storage (EES) devices are expected to play a significant role in: (i) the grid, where storage is needed to optimize use of intermittent green energy sources and support peak load times; (ii) the transportation sector, where electric vehicles (EVs) are helping to lower automobiles' reliance on fossil fuels; and (iii) personal electronics. A challenge for EES devices is optimizing design for both energy density, i.e., how much energy a device can store, and power density, i.e., how fast that energy can be delivered.[2] Batteries, which store energy through electrochemical conversions, are well-known EES devices which tend to have high energy density while suffering from relatively low power densities, translating to relatively long charging times (e.g., relative to the time needed to fill the gas tank of a car).[3] Batteries face other challenges as well, associated with issues including safety, environmental impact of mining of rare-earth components of their electrodes, and poor recyclability.[4] The lithium-ion (Li-ion) battery, a secondary battery that can be recharged, is considered one of the great accomplishments of modern electrochemistry and materials science and has become so ubiquitous in technologies such as electrical vehicles (EVs) and personal electronics that it earned the 2019 Nobel Prize in chemistry.[2, 5] Since the Li-ion battery came to market, numerous approaches have been explored to tackle its shortcomings, including improved electrode and ionic charge carrier chemistry, as well as solid-state assemblies.[6-13]

One avenue that remains relatively underexplored is the architecture of the electrode-separator-electrode assembly that comprises the EES cell of a battery. It is worth noting that there is a difference between electrode architecture and cell assembly architecture on the nanoscale. There is a large body of work both on integrating nanomaterials into existing electrodes as well as developing higher surface area, structured electrodes, such as interdigitated or cylinder arrays, on the microscale.[2, 5, 14-16] However, while such approaches can increase surface area of a single electrode and/or the interfacial area of the assembly, these electrode-separator-electrode assemblies are still fundamentally limited to layered/stacked or cylindrical morphologies. In contrast, nanoarchitecture in the present work refers to changing the fundamental form of the cell assembly architecture from a layered or sandwiched planar two-dimensional (2D) stack into one in which the anode and cathode are three-dimensionally (3D) co-continuous interpenetrating nanonetworks with the separator in between them.

EES devices with this unique architecture enabled by purposefully structured nanoscale features, so called 3D batteries, have been discussed in the literature for about two decades.[17,18] They provide high interfacial areas or high material loading per footprint area in a device. They also enable a pathway to increasing power density via shortening the distance ions need to diffuse between electrodes without sacrificing energy density due to the high active material per volume and area. Existing work on this research topic is still limited, however. This is due to the tremendous experimental challenges realizing such architectures of multiple functional materials with nanosized interpenetrating 3D co-continuous structural features in a working device.[19,20] To the best of the authors' knowledge, the only proof-of-principle example of a working 3D battery device, as described above, to date is a block copolymer (BCP) self-assembly directed interpenetrating co-continuous cubic double gyroidal 3D battery featuring lithium-sulfur chemistry and phase dimensions of all components below 20 nm.[21] Such bottom-up approaches in principle enable rapid production of nanostructures at large scale. This proof-of-principle monolithic gyroidal EES device suffered from low actual versus theoretical capacity of only ~25%, however.

Challenges associated with generating these devices are immense. The nanoscopic and co-continuous domains of redox active anode and cathode have to be simultaneously electronically conductive to carry electrons to and away from redox active sites. Generation of a conformal and pinhole free separator with nanoscale thickness is required between anode and cathode with highly curved interfacial morphologies and large interfacial areas across macroscopic monoliths to avoid short circuits. Finally, both electrodes with nanoscopic dimensions have to be reliably contacted electronically to provide overall device functionality.

Herein, we fabricate and show proof-of-principle of a working 3D EES device with critical phase dimensions of all components on the scale of ~100 nm. This is enabled by an ultra-large molar mass (ULMM) BCP self-assembly-derived co-continuous porous carbon monolith, which serves as both anode and current collector that can be electronically contacted. Onto the surface of this carbon, a REDOX-active, conductive polymer is directly electropolymerized, which takes the role of both cathode and current collector. Through overgrowth beyond the confines of the carbon anode, the conductive polymer can be separately contacted electronically. Through subsequent discharging the carbon monolith-conductive polymer hybrid against external lithium metal down to potentials that render the conductive polymer more insulating, a solid electrolyte interphase (SEI) is then formed along the carbon-polymer interface. This SEI plays the role of an effectively in-situ generated separator of the device. After the SEI formation step, the device holds potential across its two leads and can be cycled. Following an initial charge after external lithiation, the device holds a stable open circuit potential of >3.5 V for 5 hours. While the device does exhibit capacity loss over several cycles, the initial discharge post-SEI formation is 91% of the theoretical capacity. Improvements to the initial design results in devices that demonstrate reproducible discharge capacities above the theoretical capacity of the device, which is assessed assuming the cathode only accommodated lithium ions in its redox-active pendant. The successful galvanostatic cycling of these EES devices marks, to the best of the authors' knowledge, the first all-organic materials derived 3D nano-architectured Li-ion battery device and the first EES device to use an in-situ generated SEI as the device's sole separator. Furthermore, we demonstrate that the sequence of carbon template generation, electrochemical anode deposition and subsequent in-situ SEI-based separator formation can be extended to an alternative battery form factor based on polymer membrane derived porous carbons; this device reached 79% of its theoretical capacity in its second discharge. Results suggest that a large variety of 3D co-continuous porous carbons should in principle be transformable into a 3D battery via the sequence of scalable and relatively straight-forward electrochemical processes described herein. Together with organic materials derived electrode chemistries that move away from rare-earth components, our proof-of-principle work supports the feasibility of 3D architected battery devices accessible via a plethora of different 3D porous carbon supports thereby democratizing EES technology around the world.

**Materials and Methods**

**Materials**

Tetrahydrofuran (THF), n-butyl lithium (n-BuLi), sec-butyl lithium (s-BuLi), 1,1-diphenylethlyene (DPE), styrene, 2-dimethylaminoethyl methacrylate (DMAEMA), methanol, ethyl acetate, phenol, sodium hydroxide, formaldehyde, *p*-toluenesulfonic acid, acetonitrile, propylene carbonate (PC) (99%), lithium perchlorate, 3,4-ethylenedioxythiophene (EDOT) (98%), 3-bromo-1,2-propane diol (97%), 3,4-methoxythiophene, toluene, hexanes, 2-carboxylic acid anthraquinone, triethylamine, dimethylformamide (DMF), petroleum ether, 1,3-dioxolane (DOL), 1,2-dimethylethane (DME), lithium bis(trifluoromethanesulfonyl)imide (LiTFSI), lithium nitrate ($LiNO_3$), vinylene carbonate (VC) (97%), fluoroethylene carbonate (FEC) (99%), silica gel, and silver perchlorate ($AgClO_4$) were purchased from Millipore Sigma. Lithium metal was purchased from MTI Corp. Titanium wire and silver wire (>99%) were purchased from Beantown Chemical. Poly(4-vinylpyridine) (P4VP) were purchased from Sigma Aldrich and used as received. 0.45 μm average pore size nylon membranes were purchased from Sterlitech Corporation.

**BCP synthesis**

An ULMM poly(styrene-*block*-2-dimethylaminoethyl methacrylate) diblock copolymer (SA) was synthesized via sequential anionic living polymerization following known procedures.[22] Briefly, tetrahydrofuran (THF) was dried over n-butyllithium (n-BuLi) and diphenylethylene (DPE) and distilled through a Schlenk line into a 2-L reactor. Subsequently, styrene was cleaned over calcium hydride and distilled in a custom apparatus. After cooling the reactor to -78 °C, sec-butyllithium (s-BuLi) was added to the reactor, followed by the clean styrene monomer. The polystyrene was allowed to react for 1-2 hours. 2-Dimethylaminoethyl methacrylate (DMAEMA) was cleaned over triethyl-aluminum and trioctyl-aluminum and distilled in a custom apparatus. Distilled DPE was added to the living polystyrene reactor to DPE-cap the living end, followed by the addition of the clean DMAEMA monomer. The mixture was allowed to react for 1 h before quenching with degassed methanol. The reaction was carried out under nitrogen atmosphere.

**Resols synthesis**

Resols were synthesized following known procedures.[23] Briefly, a desired amount of phenol (typically 0.05-0.1 mol) was added to a round bottom flask equipped with a reflux condenser, which was put in a 45°C water bath under stirring. Once the phenol melted, 20 wt% aqueous sodium hydroxide (NaOH) was added (10:1 molar ratio of phenol to NaOH), and the mixture was allowed to stir for 10 minutes. Formalin solution was then added dropwise to the sodium phenoxide solution until a 1:2 molar ratio of phenol to formaldehyde was achieved, and the temperature of the bath was raised to 70 °C. After refluxing at 70 °C for an hour, the solution was allowed to cool, neutralized with *p*-toluene sulfonic acid, and freeze dried. The dried product was dissolved in THF to precipitate out the salt, filtered, and freeze dried again. This product was dissolved to desired concentration in THF.

**Co-assembly and pyrolysis of monolithic carbon films (anode fabrication)**

Ultra-large porous carbons (ULPCs) were synthesized from a procedure adapted from earlier studies.[22,24] The carbons were co-assembled via evaporation induced self-assembly (EISA) by casting a 2 wt% solution of BCP:resols at a mass ratio of 1:2 in a solvent mixture of 8:2 (by weight) THF:ethyl acetate into a poly(tetrafluoroethylene) dish. The solutions in the dish were placed on a hot plate within a dry box with a nitrogen environment, covered with a glass dome, and allowed to evaporate with the hot plate set to 50 °C until there was no more condensation left on the dome (typically > 12h). The resultant films were then crosslinked by increasing the temperature of the hot plate to 120 °C for another 12 h while maintaining the nitrogen atmosphere. The hybrids were then cut into the desired shape and size and carbonized under nitrogen. The pyrolysis consisted of a 1 °C per minute ramp up to 600 °C, followed by a 3 hour hold at 600 °C, and then a 5 °C per minute ramp to 1100 °C, which was held for 1 h.

**Cathode monomer synthesis**

The monomer, (2,3-dihydrothieno[3,4-b][1,4]dioxin-2-yl)methyl 9,10-dioxo-9,10-dihydroanthracene-2-carboxylate (AQEDOT) was synthesized in a two-step reaction. First,

EDOT with a brominated pendant, 2-(bromomethyl)-2,3-dihydrothieno[3,4-b][1,4]dioxine (Br-EDOT), was synthesized following known procedures.[25] Briefly, 5 g of 3-bromo-1,2-propanediol was reacted with 5 g of 3,4-dimethoxythiophene and 1.5 g of *p*-toluene sulfonic acid in 80 mL of toluene at 100 °C for 20 h. The product was filtered to remove precipitated black solids, and the remaining liquid product was diluted with ~40 mL dichloromethane. The organic layer was washed with water, then brine, and extracted with a separatory funnel. The product was then dried from the organic layer using a roto-evaporator and purified over a silica column using 3:1 hexanes:dichloromethane by volume. Next, the resultant brominated-EDOT (Br-EDOT) product was reacted with 2-carboxylic acid anthraquinone.[26] The two reactants were added at a 1:1 molar ratio in DMF (~10 wt% solution). Triethylamine was then added to the solution (1.5:1:1 molar ratio of triethylamine:2-carboxylic acid anthraquinone: Br-EDOT) and the solution was held stirring at 85 °C for 14 h. The product was poured into water and extracted with ethyl acetate, which was washed with brine; the organic component was then separated in a silica column over 5:1 v/v petroleum ether: ethyl acetate and collected. The product was characterized with proton NMR on a Bruker AV 500 MHz spectrometer, and peaks were compared against known values from literature (Figure S1).[25, 26]

**Cathode electropolymerization**

A polymerizing electrolyte of 0.1 M lithium perchlorate (LiClO$_4$) and 0.01 M AQEDOT in 1:1 v/v acetonitrile (MeCN) and propylene carbonate (PC) was prepared. An ULPC monolith was weighed and then attached to a titanium wire with a carbon paste made of carbon black and a poly(vinylidene fluoride) (PVDF) binder to act as a working electrode (WE). The wire was connected along the middle of one of the main faces of the carbon monolith, with the conductive paste covering the entire face. The wire and monolith were then orientated so that the remaining uncovered face was flat and facing upwards. Insulating epoxy was used to cover the downwards-pointing face of the ULPC monolith that was in contact with the carbon paste and wire, resulting in only the upwards-pointing uncovered face and cross-section remaining accessible. After the epoxy was fully cured (>90 min at 80 °C in air atmosphere), the assembled WE, comprised of the ULPC monolith, wire, conducting paste, and insulating epoxy, was weighed and the mass was recorded. The WE was then put under vacuum for >12 h and was subsequently immersed in the electrolyte solution for >1 h. This WE, now impregnated with electrolyte, was then suspended in an electrochemical reactor purged with argon, and drops of electrolyte were added to the WE's upward-facing surface until a droplet with surface tension was formed. A silver wire serving as both a counter electrode (CE) and a pseudo-reference electrode (RE) was carefully lowered into the set up until it contacted the droplet.

Electropolymerization was carried out on a Metrohm Autolab PGSTAT204 using a custom-made piece of glassware made to accommodate 3-electrodes and an argon purge (Figure S2). Cyclic voltammetric (CV) deposition was then carried out for 5 cycles between 2.5 V and -0.6 V at 10 mV/s. This was repeated 1-2 times depending on the size of the carbon. A pulsed potential deposition method was also explored, where 2.5 V was pulsed for 1 second to polymerize the AQEDOT. The set-up was then allowed to rest for 5 seconds to enable diffusion of monomer back into the pores, and, finally, a 5 second pulse at the reducing potential of -2.5 V was applied before repeating the process 250 times. After washing the electrolyte off with acetonitrile and IPA, followed by drying the sample for at least 8 h at 80 °C under vacuum, it was weighed again to determine the weight of the polymer deposited.

**Current collector electropolymerization**

To add a pure poly(3,4-ethylene dioxythiophene) (PEDOT) current collector, the WE,

now with poly((2,3-dihydrothieno[3,4-b][1,4]dioxin-2-yl)methyl 9,10-dioxo-9,10-dihydroanthracene-2-carboxylate) PAQEDOT in it, was immersed into electrolyte of 0.02 M EDOT and 0.1 M LiClO$_4$ in a 3-electrode cell glassware set up with a Ag/Ag$^+$ RE (silver wire with 0.1 M LiClO$_4$ and 0.05 M AgClO$_4$), a platinum CE, and purged with argon gas (Figure S2). Using this set up and a Metrohm Autolab PGSTAT204 potentiostat, the WE was cycled between -0.6 V and 2.5 V vs. Ag/Ag$^+$ at 10 mV/s for 5 cycles to deposit EDOT primarily on the exposed surface of the monolith. After electropolymerization, the WE was rinsed in acetonitrile and IPA, and then dried for at least 8 h under vacuum at 80 °C and weighed. Afterwards, a second contact was made to the face with the PEDOT current collector (Figure S3) with carbon paste over the entire exposed surface, being careful to leave the cross-section uncovered, and finally capping that connection with insulating epoxy.

**Electrochemical processing**

As of this step, the device was physically fabricated, but had no lithium, nor any separator between the carbon anode and polymer cathode. To complete the EES device, electrochemical processing was required. The device was set up in a septa-capped vial, along with another wire and alligator clip, with the wires penetrating through the septa. In a glovebox, an electrolyte of 1 M LITFSI in 1:1 v/v DOL:DME with 2 wt% LiNO$_3$ and 1.5% VC was prepared. Then, under argon, a lithium chip was affixed to the alligator clip and the electrolyte was added to the septa-capped vial such that the monolithic device and the lithium chip were immersed in the electrolyte. First, the carbon was discharged to 0.01 V vs the external lithium and then allowed to rest for 10 minutes; this was repeated four more times to generate SEI. This was followed by a discharge to low potentials between -0.3 V and 0 V to find a plateau at which lithium would be plated and more electrolyte decomposition would take place. This low potential plateau was maintained until >132 mAh/g, the theoretical capacity of AQEDOT, was reached. The carbon was then charged to 4.0 V to oxidize it and strip the plated lithium.

Next, to lithiate the PAQEDOT, the cathode portion of the monolith was discharged to 1.0 V vs lithium. After this step, with the anode side in an oxidized state and the cathode in a reduced state, the device was in a discharged state. At this point, the device was charged in the electrolyte solution, attaching both connections to each of the anode and cathode, and no longer using the external lithium chip. After the SEI forming step, if the SEI would indeed act as a separator, the device should hold a potential; if there would be a short, no potential would be held, as each electrode would be a conductor in contact with each other. The device was charged to 4.2 V and then held at open circuit for 5 h to monitor the open circuit potential (OCP).

**Polymer membrane-based devices**

To demonstrate the generalizability of the method to other co-continuously porous carbon sources, the processing steps described above, i.e., cathode electropolymerization, additional current collector electropolymerization, SEI formation, and further electrochemical processing were applied to a carbon anode derived from a soft polymer membrane template. Synthesis of the carbon membrane anode was adapted from a procedure detailed elsewhere.[27] Briefly, THF was evaporated off of the resols solution using vacuum, and the solvent was swapped to DMF with a final concentration to 2 wt % resols in DMF. P4VP and DMF were then added to the solution to yield a 2 wt % resols and 2 wt % P4VP solution in DMF, which was used for infiltration of 0.45 μm average pore size nylon membranes (Sterlitech). To that end, the nylon membrane were cut into 1 cm$^2$ area squares. These square pieces were placed on a PTFE dish and heated on a hot plate to 90 °C. Then 400 μL of the resols/P4VP/DMF solution was deposited on the square pieces, in 100 μL increments, ensuring that the solution did not flow over the area of the nylon membrane. After the solution was evaporated, the temperature was raised to 120 °C to crosslink the resols for >12 h. These films were then pyrolyzed using the same way and using the same heating profile as detailed above.

The resultant carbon membranes were then subjected to the same workflow as detailed above – mounting on a wire, being backfilled with electropolymerized cathode, having a PEDOT current collector deposited on the surface, connection of the cathode current collector to a second external lead, and electrochemical processing for self-separation – to yield alternative 3D self-separating all-organic batteries. These devices were also externally lithiated, tested for OCP, and cycled.

**Structural characterization**

SEM imaging and energy dispersive X-ray (EDX) imaging were performed on a Zeiss Gemini 500 SEM. Secondary electron images were taken at a 3 kV accelerating voltage and EDX imaging was performed employing AZtec software at an accelerating voltage of 10 kV.

**Cycling and electrochemical characterization**

If the OCP was stable at over 3 V for the duration of the hold, devices were considered viable for cycling. To that end, a device was lifted out of the electrolyte to run in the solid state. It was then cycled with a desired constant current on an MTI 8-Channel battery tester. After cycling the device was subjected to electrochemical impedance spectroscopy (EIS).

Cycling data was taken on an MTI 8-channel battery tester using TC 5.3 software. EIS was performed over a frequency range of 1 Hz to 1 MHz on a Biologic VMP3 multi-channel potentiostat with the cathode as the working electrode and anode as the counter electrode. EIS data were recorded and fits were performed on EC-Lab V11.50 software, using a proposed equivalent circuit and a Randomize (10000 iterations) + Simplex (1000 Monte-Carlo simulations) method for the fit.

**Results and Discussion**

*General approach and self-assembled porous carbon monolith formation.* This work describes the realization of all-organic materials derived full-cell EES devices with 3D interpenetrating electrodes. It also introduces the concept of in situ generation of a separator between anode and cathode. To that end, the cathodes are directly electropolymerized onto 3D co-continuous porous carbon monoliths that act as the anode and both electrode materials are separately contacted. The separators are subsequently formed in situ by discharging the anode materials against external lithium in an electrolyte at potentials where the cathode materials are more insulating. Such devices are unique from a conventional lithium-ion battery in the following ways: (i) they exhibit an interpenetrating co-continuous morphology of the electrode-separator-electrode architecture; (ii) the critical thickness length scale of the electrode-separator-electrode assembly is only of order 100 nm; (iii) the battery electrodes are generated using all-organic materials; (iv) the battery separator/electrolyte material is produced by using the electronically insulating but ionically conducting SEI itself to separate the electrodes. Each fabrication step required deliberate design to enable all the features of these proof-of-principle 3D devices.

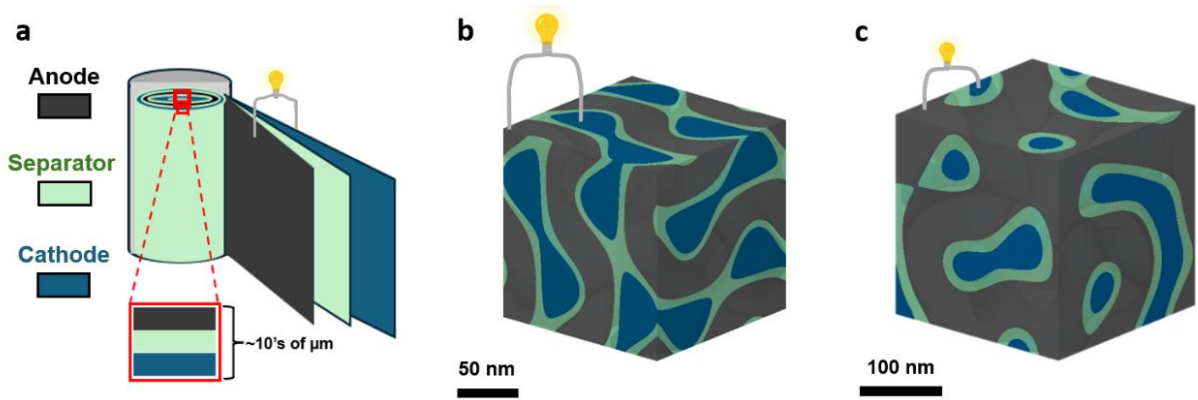

***Figure 1.*** *Comparison of battery architectures. (a) Conventional battery with 2D stacks of planar sheets of anode, separator, and cathode (electrodes applied to a current collectors; not shown as separate entities), with thickness on the scale of tens of microns. (b) First cubic co-continuous double gyroidal 3D battery cell from reference 21 with interpenetrating electrodes separated by a solid electrolyte electropolymerized onto a porous carbon scaffold. (c) Next-generation 3D battery composed of interpenetrating ultra-large porous carbon anode and redox-active, conductive polymer cathode with an in-situ-grown solid electrolyte interphase (SEI) as separator. Please note that although the cathode and separator domains seem isolated from each other, like the anode domain they also form co-continuous networks (e.g., see images in figure 3).*

The differences between conventional and 3D battery designs on the macro and micro/mesoscale are highlighted in **Figure 1**. While a cylindrical cell is one of several form factors of conventional battery devices, the trait that all form factors of current batteries share, from coin cells to pouch cells to prismatic cells, is that the assemblies are, in principle, stacked 2D planar sheets of electrodes and separator (**Figure 1a**). Typically, these devices are fabricated by coating slurries containing solvent, a polymer binder, particulate active material, and a conductive additive onto metal current collectors. These electrodes are then layered with a separator in between them, and the assembly is packaged into the desired form factor (e.g., via rolling). While the electrodes may be modified with microstructured surfaces or interdigitation,[16] the dimensionality of the path from one electrode sheet, through the separator, to the other does not change. This is contrasted with two types of 3D EES devices in **Figure 1b** and **c**. Both are fabricated from co-continuous porous carbon monoliths (dark colored anode material) and show 3D interpenetrating electrodes (cathode materials in dark blue) divided by a separator (light green domains), exhibiting a stark contrast to conventional designs in both length scale and morphology of the cell assembly architecture. While the 3D device in **Figure 1b** described in an earlier study is constructed from a BCP self-assembly directed interpenetrating co-continuous cubic gyroidal equilibrium morphology,[21] the one in **Figure 1c** shows a non-periodic structure directed from ULMM BCP non-equilibrium self-assembly resulting in pore sizes about twice that of the gyroidal material (~100 nm vs. ~40 nm pores in the gyroidal assembly). Please note that while not evident from Figure 1c, anode, separator, and cathode domains are co-continuous throughout the entire monoliths. EES devices from such non-equilibrium structures are the focus of this paper and, with a similarly thin separator generated in-situ, exhibit substantially higher theoretical capacity than prior devices.

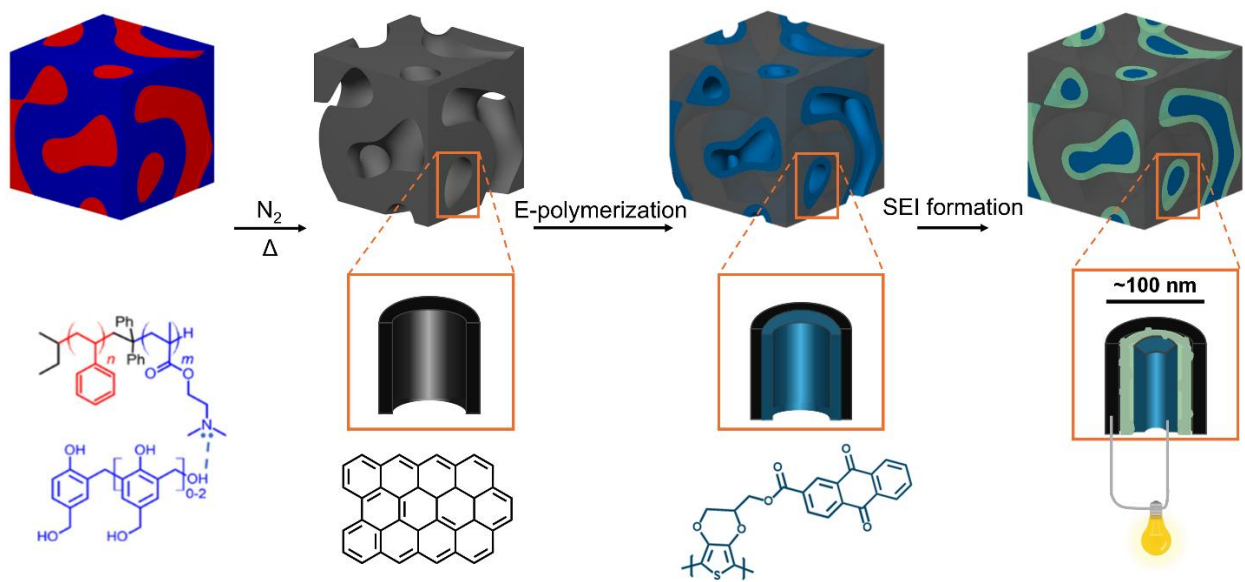

*Figure 2.* Schematic for the fabrication of self-separating all organic 3D battery and molecular structure of constituting organic structures/framework. From left to right, the device starts as a co-assembled hybrid from ultra-large molar mass (ULMM) BCP plus resols, which is then pyrolyzed under nitrogen into a large pore (~100 nm diameter) carbon. Counterintuitively, this carbon is then first used as a working electrode to electro-deposit/polymerize the cathode polymer. The carbon is then reduced and lithiated via discharge against lithium metal in electrolyte to generate a solid-electrolyte interphase (SEI) at the carbon anode-polymer cathode interface at potentials where the polymer cathode is insulating to generate the final device structure.

The approach to generate such monolithic devices with mesoscopic, co-continuous interpenetrating electrodes is shown in **Figure 2**. In the first step, as described in detail elsewhere,[22,24] an ULMM poly(styrene-*block*-2-dimethylaminoethyl methacrylate) (SA) BCP with molar mass of 913 ± 83 kg/mole (determined by Zimm plots) and 80 vol% poly(styrene) (S) block was used to structure direct phenol-formaldehyde resols to generate co-continuous organic-organic hybrid materials via evaporation induced self-assembly (EISA) (Methods). Resulting hybrid films were then cut into desired rectangular shapes using a razor blade. These pieces were subsequently carbonized via pyrolysis to generate the rectangular co-continuous carbon anode monoliths that simultaneously serve as current collectors used as a starting point for EES device formation (Methods). To that end, after contacting with a wire, employing these carbon monoliths as electrodes, the pore surfaces were subsequently decorated through electropolymerization with a polymer that was simultaneously suitable as a current collector and cathode material. This polymer was poly((2,3-dihydrothieno[3,4-b][1,4]dioxin-2-yl)methyl 9,10-dioxo-9,10-dihydroanthracene-2-carboxylate), i.e., PAQEDOT (Methods). This resulted in two interpenetrating networks of conductive electrode materials in direct contact to each other. Lastly, after contacting the cathode with a second wire and to complete the electrode-separator-electrode assemblies of this type of 3D co-continuous EES cells, an SEI was generated in-situ at the interfaces of the two electrodes serving as a separator.

As-assembled and resulting porous carbon anode monoliths of the EES cell cut into rectangular shapes at the as-made stage are shown together with scanning electron microscopy (SEM) images of cross sections of the carbon material in **Figure 3a-d**. Carbon is an ideal material candidate to enable this design, as it can serve as an anode, a current collector, and a robust 3D porous scaffold. The co-continuous pore structure of the carbon evident from the SEM images played a key role in enabling the fabrication of the devices. First, co-continuous morphologies tend to shrink isotropically. This allowed the hybrid films to maintain their macroscopic monolithic shape and structural fidelity upon conversion to carbon via pyrolysis (**Figure 3a**). The co-continuous pore space (**Figure 3b-d**) present in the carbon anodes enabled subsequent filling with separator and active materials (vide infra) to allow for very high device interfacial areas in a small areal footprint. Previous work showed 3D EES devices with lithium-sulfur chemistry enabled by BCP self-assembly-derived carbon anodes with double gyroidal morphology and ~ 40 nm pores.[21] In the present work, while a BCP structure-directing agent was also chosen, larger pore sizes ~100 nm were targeted. To that end, an ULMM BCP, ~1000 kg/mol, was used. While gyroidal structures were envisioned, the ULMM of the polymer instead resulted in a structure that was not periodically ordered, likely the result of polymer entanglements and associated long structural relaxation times. Even though no periodic order was achieved, the resulting structures were still co-continuous in their carbon and pore spaces, respectively. The generated pores had an average size of approximately 90 nm (**Figure 3d**), close to the desired target. The pore size of ~100 nm is notably a difficult length scale to reach for co-continuous structures using conventional linear BCP self-assembly.

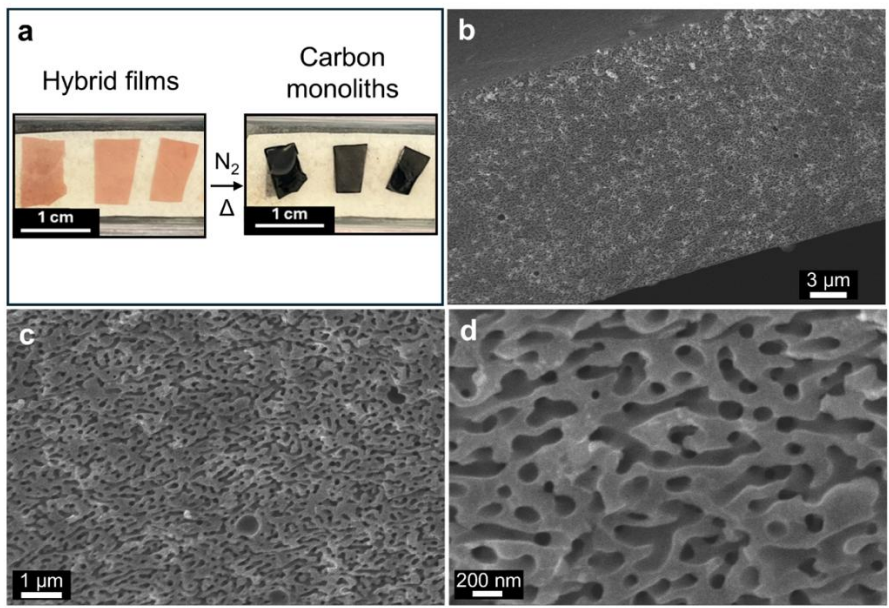

*Figure 3. Fabrication and structure of porous carbon anode materials. (a) Conversion of co-assembled hybrid films to monolithic carbons on the macroscale. (b-d) SEM cross-sectional imaging of porous carbon anode monolith progressively zooming in to the local pore network.*

*Elelctropolymerization of PAQEDOT based cathode/current collector directly onto the carbon framework and its contacting with a second metal wire lead.* The next step after fabricating ultra-large porous carbon (ULPC) anodes and contacted by a wire (Methods) was, counterintuitively, adding the cathode material. Requirements for the cathode material included being conductive and redox-active (ideally, in a single phase), and processable in a way that would allow for efficient back-filling into the pore space, an issue that had plagued the original gyroidal devices.[21] For example, micron-sized particles of transition metal oxide, such as those used in conventional Li-ion cathode slurries, would not work for the 3D cell targeted here, since such sizes would be too large to homogeneously infiltrate the pores across a macrocopic monolith. To this end, a monomer, AQEDOT, was chosen since it could be electropolymerized into a conductive, redox-active polymer from an electrolyte solution. Organic cathode materials have drawn attention in battery research in recent years due to their potential sustainability, processibility, enhanced ion mobility, affordability, and safety advantages over conventional Li-ion inorganic electrode materials.[26, 28-32]

Figure 4a shows a schematic set-up of the electropolymerization required to generate PAQEDOT on the pore surfaces within the ULPCs. A wire was afixed to the anode with carbon paste to be used as a working electrode (WE). We introduced a bend at the end of the wire so that a polished surface of the carbon monlith was face up. Insulating epoxy was added to the bottom face of the carbon, where the wire-carbon connection was previously made with carbon paste, to add structural integrity and electronically isolate the wire from anything but the carbon. This WE was subjected to vacuum for > 12 h and then soaked in electrolyte containing the monomer for at least 1 h. An additional drop of electrolyte was placed on the upward-facing surface of the carbon WE as shown in Figure 4a. A silver wire, serving as both counter electrode (CE) and pseudorefernce electrode (p-RE), was then contacted with the surface of the electrolyte droplet. This electrode set up with a single silver wire serving as CE and p-RE has previously been used to polymerize EDOT derivatives with organic pendants.[33] Here, the "reactor in a droplet" configuration was designed with the intent of forcing as much polymer deposition into the pores as possible. This was achieved by keeping monomer concentration within the pores and on the surface of the WE similar. This is in contrast to immersing the WE into a polymerizing solution and exposing the surface to a large bath of polymerizing electrolyte. The solvent for the polymerization electrolyte was chosen as a 1:1 v/v mixture of acetonitrile to propylene carbonate (PC). This mixture has been described in the literature suggesting that while PC has higher PEDOT oligimer solubility and results in smoother films of PEDOT, it is too good of a solvent for deposition of the polymer into porous electrodes.[34, 35] Pure acetonitrile would in principle result in more polymer deposited, but also ran the risk of depositing more polymer onto the WE surface. Therefore a mix of these two solvents was chosen. It is likely that further optimization of the ratio of acetonitrile to PC could tailor the solubility properties of the electrolyte to favor deposition in the pores over the surface even more but was not pursued in this study.

The monomer was initially electropolymerized using a cyclic voltametric deposition method, sweeping between -0.6 amd 2.5 V vs Ag at a speed of 10 mV/s (Figure 4d). This deposition method resulted in a sample with a 3:1 mass ratio of carbon to PAQEDOT, which was further processed into a device. After electrodeposition, there was noticible polymer in the pores of the carbon (Figure 4e) as compared to a bare porous carbon anode (Figure 4b). While such images suggested that pores were not homogeneously filled with polymer on the pore level, EDX imaging of sulfur, an element only present in the PAQEDOT, across the cross-section of the monolith showed relatively homogeneous sulfur signal throughout on the cell level (compare

Figures 4c and 4f).

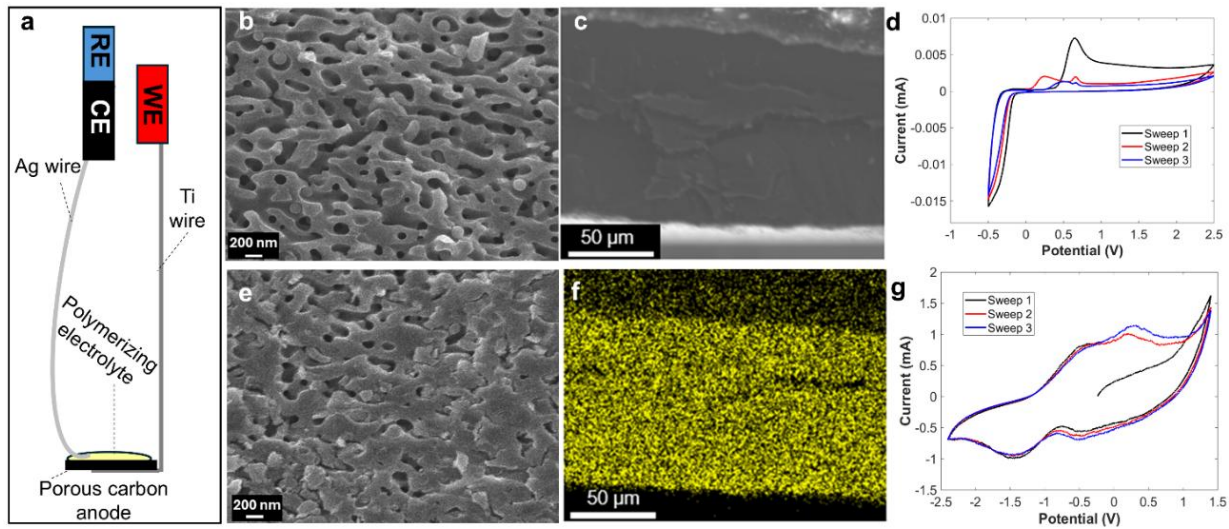

*Figure 4.* Integration of polymeric cathode material. (a) Schematic of the set up for electropolymerization with droplet of electrolyte on flat anode surface and silver wire acting as counter electrode and pseudoreference electrode. (b) SEM cross-sectional image of uncoated ULPC anode pores. (c) Lower magnification SEM image of post-electropolymerization monolith cross section. (d) Cyclic voltammogram of the electropolymerization of AQEDOT. (e) SEM cross-sectional image of pores after P(AQEDOT) deposition. (f) Lower magnification SEM image of monolith cross section with EDX trace of sulfur (from PAQEDOT) in yellow. (g) Cyclic voltammogram of PAQEDOT deposited on UPLC in supporting electrolyte of 0.1 M LiClO$_4$ in 1:1 v/v MeCN:PC.

Cyclic voltammetry (CV) taken of a PAQEDOT film deposited on the ULPC WE in supporting electrolyte (0.1 LiClO$_4$ in 1:1 MeCN:PC) showed a reduction peak at ~ -1.5 V vs Ag/Ag$^+$ and an oxidation peak ~ -0.4 V vs Ag/Ag$^+$ (Figure 4g). These two lower peaks are expected to be associated with the anthraquinone pendant in PAQEDOT (Figure S4), with the smaller peaks associated with each individual carbonyl group washed out into a single peak due to the mass transport limitations in a thick polymer film. The reduction potential of an Ag/Ag$^+$ reference electrode in solution will vary with organic solvent and electrolyte concentration.[36] Assuming a Ag/Ag$^+$ reduction potential of 0.8 V vs SHE and a Li/Li$^+$ potential of -3.04 V vs SHE (standard hydrogen electrode), and using our reduction plateau of -1.5 V against Ag/Ag$^+$, we can expect a reduction plateau from PAQEDOT vs. lithium around 2.3 V.[37] The peaks located at ~ -0.5 V (reduction) and 0.3 V (oxidation) vs Ag/Ag$^+$ are expected to be associated with the reduction and oxidation of thiophene in the PAQEDOT backbone, respectively.[34]

While the PAQEDOT is both conductive and REDOX-active in a single phase, the polymerization did not generate a thick enough polymer film on the monoliths outer surfaces to serve as current collector to which a lead could be attached without also touching carbon. To ameliorate this situation, a film of pure PEDOT was electropolymerized via CV at 20 mV/s in a more traditional 3-electrode set up with a platinum counter electrode and Ag/Ag$^+$ reference electrode immersed in 0.02 M EDOT and 0.1 M LiClO$_4$ in MeCN (Figure S2, Figure S3). This subsequent deposition method prioritized film generation on the surface of the carbons, since the monomer concentration in solution at the surface of the monolith is more quickly replenished than in the torturous internal pore system. A second metal wire lead was subsequently contacted to this PEDOT current collector with carbon paste.

As previously emphasized, the design of adding the cathode material directly in contact with the anode material at first sight seems counterintuitive to the goal of generating an EES device that requires electronic separation between anode and cathode phases without short circuits. However, using the carbon anode as the WE does allow for generation of cathode polymer films at every interface between the carbon and polymerizing electrolyte, ideally filling internal surface areas with the cathode material. It is this order of processing that led the authors to refer to this assembly as a "self-separating" cell, as there initially is no physical separator added during fabrication of the two electrodes of the physical devices. This non-sequential manufacturing approach borrows concepts from so-called "anodeless," or anode-free, batteries, which have been recently described in battery research.[38, 39] In anodeless batteries, the cells are assembled with no anode material, but rather just a current collector with space for lithium to be plated and stripped, acting as an anode material that is generated during the electrochemical process of cycling the device. Likewise, in the design of our 3D EES device, we intended to generate our separator in situ during electrochemical processes, as described in the following section.

*In-situ generation of a SEI layer for electrode separation.* With the addition of an insulating epoxy to cover the leads and improve robustness of the devices, samples were ready for in-situ generation of an SEI layer for electrode separation. To that end, the devices were enclosed in a vial with electrolyte (1 M LiTFSI, 2 wt% $LiNO_3$, and 2 wt% VC in 1:1 DOL:DME) and an external lithium chip. The electrolyte composition was chosen to include a salt with high potential stability and fluorine content, as well as additives known to help generate robust, stable SEI with both organic and inorganic components.[40, 41] Figure 5 outlines the electrochemical steps to go from assembled and contacted electrodes to completed EES devices.

The initial discharge of the carbon vs the external lithium was to generate SEI along the carbon surface (Figure 5a). Critically, SEI tends to form on carbon below ~1.0 V vs Li.[42, 43] At these potentials, de-doping of the polymer backbone tends to make PEDOT and its derivatives more insulating.[44] Therefore, the goal of the design was to form the SEI between the conductive carbon pore wall and the insulating polymer in the pore. After initial discharge, a lower potential discharge was carried out (Figure 5b). This second discharge was performed at a potential expected to plate lithium, which should lead to more electrolyte decomposition and SEI formation, as well as lift polymer off of the carbon surface. Subsequently, the lithium was stripped from the carbon as it was charged to 4 V vs lithium.

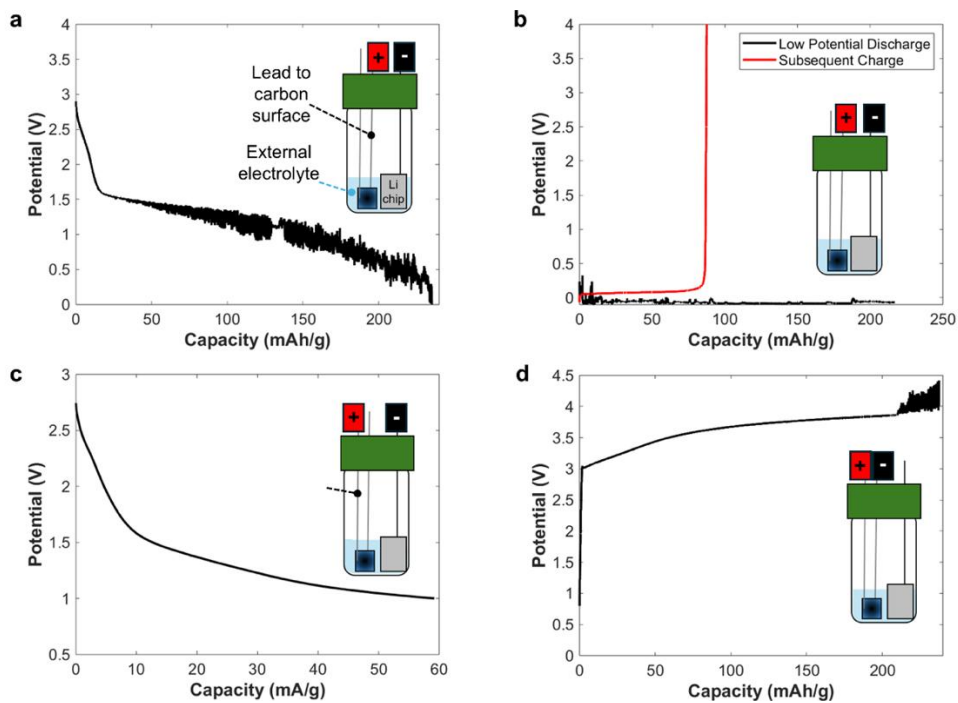

*Figure 5.* Electrochemical processing from connected anode and cathode materials to electrochemical energy storage (EES) devices that can separate and hold charge. (a) Initial discharge of carbon versus an external lithium chip in electrolyte. (b) Subsequent discharge at a lower potential followed by oxidation of the carbon. (c) Lithiation and reduction of PAQEDOT versus external lithium chip, leading to devices in the discharged state. (d) Charging of devices in electrolyte. The capacities of (a) and (b) are normalized by the mass of carbon and the capacity of (c) is normalized by the mass of PAQEDOT.

The next step was to "lithiate" the PAQEDOT via reduction (discharge) vs the external lithium chip (Figure 5c). Since the PAQEDOT was in contact with the carbon monolith during its initial discharge prior to SEI formation, it was likely already in a partially reduced state. This step was to ensure this was indeed the case. With the carbon in an oxidized state and the PAQEDOT in a reduced state, one would expect the cell to be in a discharged state. The following step was to charge the devices without use of the external lithium (Figure 5d). The devices were charged while in the electrolyte to replace any de-doped counter ions on the PEDOT backbone lost during its reduction. Of note, when the leads were attached to the carbon anode and PAQEDOT cathode, there was an open circuit potential (OCP) of ~ 0.8 V in the discharged state. This already supported the idea that the electrodes had successfully been separated electronically, as conductive carbon in contact with conductive polymer would be a short and not be able to separate potential.

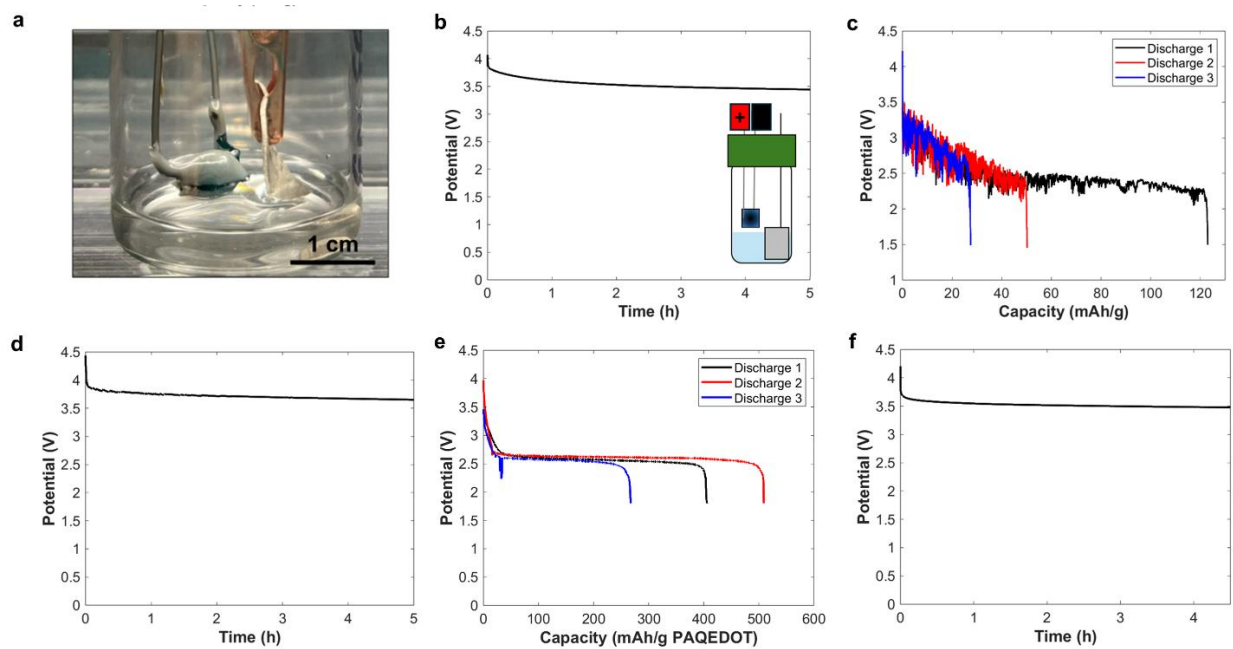

*Figure 6.* Electrochemical performance of the devices. (a) Picture of set up showing a device on the left and alligator clip holding lithium chip for external lithiation on the right. (b) Resting open circuit potential after initial charge, showing a stable separation of charge. (c) Initial discharges, with constant current charging to 4.2 V after each discharge. (d) Open circuit potential after cycling. (e) Initial three discharge cycles of further improved BCP-directed carbon-based 3D EES device after charging to ~4.1 V between each discharge. (f) OCP hold after cycling the device, demonstrating no short circuits forming during operation.

Figure 6a shows a photograph of the actual device set up. To test the promise of the device to separate and hold charge, it was left at an open circuit for 5 hours after charging (Figure 6b). After 5 hours, the EES device showed an OCP of over 3.5 V, successfully demonstrating prolonged separated charge, indicating that these electrochemical processes did indeed result in the electronic separation of the two electrodes. After confirmation of the stable charged OCP, the device was cycled. Of note, prior to cycling the devices were lifted out of the electrolyte bath so that they could be run in the solid state. Figure 6c shows the discharge capacity curves of a device discharged at a constant current of 35 mA/g PAQEDOT and charged to 4.2 V between each discharge at the same constant current. Discharge 1 was 120 mAh/g, 91% of the theoretical capacity, however, charges 2 and 3 showed subsequent capacity fade. Of note, the discharge plateau was about 2.4 V versus lithium, which corroborates the position of the reduction peak in Figure 4g. Further, literature supports a reduction potential near this value for this cathode material.[26] After three discharges, the discharge capacity was 20.8% of the 132 mAh/g theoretical capacity of PAQEDOT. While there are questions in the field of whether nanomaterials in EES devices lean more towards battery-like or pseudocapacitor-like behavior, the plateaus with steep drop offs in Figure 6c indicated battery-like behavior.[45] There was instability in the potential during both discharge and charge, possibly due to uneven current flux throughout the monolith, which possibly resulted in the 4.2 V charge point being reached before the charge capacity reached its maximum; this possible incomplete charge at 4.2 V could have also been due to overpotential. In either case, the discharge capacity fade could have been due to some issues associated with fully charging the battery before reaching 4.2 V. After the final charge, the cell was held at an open circuit to see if the OCP was still maintained, indicating the separator maintaining fidelity (Figure 6d). Once again, after 5 hours, the OCP was stable at over 3.5 V.

     Finally, Electrochemical impedance spectroscopy (EIS) was run on the device after cycling, with the resultant Nyquist plot showing a shape that could be fit to an equivalent circuit with elements indicative of an EES device (Figure S5). The equivalent circuit, shown in Figure S5b, was an electrode in series with a separator and a second electrode. In general, the curve showed higher impedance than a typical battery, but it was similar in impedance to the 3D nanoarchitected EES device described previously.[21] While it is not uncommon for solid state

battery cells to exhibit impedance values on the scale of kilohms (kΩ), the rather high impedance values of the 3D battery with, in principle, short path lengths could be due to issues such as unfavorable contact between interfaces in the device, possibly due to volumetric changes in the electrodes during cycling.[46] The shape of a semi-circle in Figure S5a, while imperfect, followed by a tail at an angle slightly above 45° associated with semi-infinite diffusion, corresponding to simultaneous processes of ionic resistance, charge transfer resistance, double layer capacity, and Warburg impedance, i.e., impedance caused by diffusion, indicates battery-like behavior.[47, 48] In future studies, it may be possible to design a thin film battery with in-situ generated SEI and probe with EIS in order to deconvolute impedance measurements of the full cell.[46]

*Further BCP derived device optimization.* An iteration on the processing of the BCP-based devices was carried out in an attempt to increase the amount of active cathode mass per unit carbon anode mass and thereby to enhance performance of the BCP-based devices. A detailed discussion can be found in the Supplemental Information (SI). Briefly, the electropolymerization method of the cathode was switched to a pulsed deposition (Figure S6), which yielded a 3:2 carbon:PAQEDOT mass ratio, compared to 3:1 of the original deposition method into the BCP-derived carbon. From there, the same current collector deposition and electrochemical processing (Methods) were carried out (Figure S7) to finalize the devices.

Figure 6e shows the cycling behavior of a battery derived in this way for the first three cycles at a constant current of 22.5 mA/g PAQEDOT, while Figure 6f shows a stable OCP after cycling. The devices generated in this iteration had very high capacities in early cycles (1-3), still exceeding 150% of theoretical capacity at the third discharge, but once again exhibited significant fading after the third cycle (Figure 6e, Figure S8). The devices were also able to maintain OCP after cycling (Figure 6f). The high discharge capacities may be explained by the long constant current charges that hit a large plateau before their cut-off potential (Figure S9). A device where only PEDOT was used as a cathode material, rather than PAQEDOT, was generated to see if PEDOT contributions from (i) the cathode backbone and (ii) the current collector could explain the capacities exceeding 100% of the theoretical capacity. While a device with PEDOT as the only cathode material indeed showed capacity, it electrochemically behaved like a capacitor and lacked the faradaic plateaus of the PAQEDOT cathode devices (Figure S10). To better understand the cyclability issues, a CV of a thin film of PAQEDOT was taken (Figure S11). Discrepancies in peak position between scans and changing peak area indicated issues with diffusion limitation and potential irreversible side reactions. These results, combined with discharges in Figure 6e indicating behavior moving beyond the theoretical capacity, support the notion that there are potential irreversible degradation reactions occurring during cycling.

*Generalization to devices from other carbon form factors.* A particularly appealing strength of the manufacturing design described so far is that, in principle, it should be generalizable to any self-supporting carbon source with a co-continuous porous morphology. To test this hypothesis, 3D cells were developed using carbons that were generated from a commercial polymer membrane as a soft template for a resols-homopolymer mixture.[27] The resultant carbons had macropores that were on the hundreds of nanometers to one micron scale (Figure 7a). These carbons were then processed into a 3D EES device using the same procedure as previously described, i.e., mounting on a wire, depositing a PAQEDOT cathode (Figure 7b), depositing a PEDOT surface layer, connecting a second lead, discharging against lithium in electrolyte to generate SEI (Figure 7c), and charging in electrolyte and testing for a stable OCP (Figure 7d). Figure 7e shows the successful cycling behavior of this generalized 3D EES device. The capacity after three discharges was about 156 mAh/g, 118% of the theoretical capacity of the PAQEDOT and higher than what was observed for the initial BCP-derived form factor (~21%). The capacity fade accelerated past the third cycle. After five cycles the discharge capacity had fallen to 32% of the theoretical capacity. The device was also tested as a full cell with cyclic voltammetry, which showed a reduction peak around 2.4 V, corroborating the discharge plateaus seen in Figure 7e (Figure S12). The demonstrated higher capacity could have been due to the larger pores allowing a more conformal deposition of more polymer in the cathode electropolymerization step, as demonstrated by a carbon to polymer mass ratio of 2:1, versus the ratio of 3:1 for the BCP-based carbon using a CV deposition method.

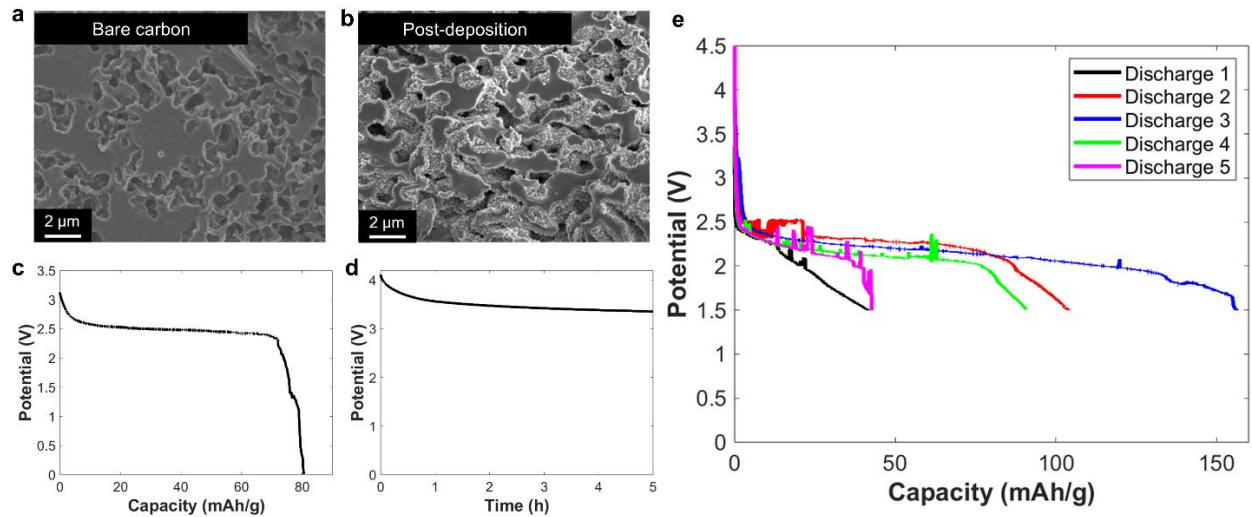

*Figure 7.* Generalization of procedure to other co-continuous carbon sources. (a) SEM of cross-sectional fracture face of membrane templated carbon before electropolymerization. (b) SEM of cross-sectional fracture face of membrane templated carbon after electropolymerization of cathode in pores. (c) Profile of SEI-generating step, discharging carbon against lithium in electrolyte. (d) 5-hour hold of OCP after initial charge of device post-SEI generation. (e) Cycling behavior of membrane-templated carbon-based EES device, showing improved performance over the initial BCP-directed carbon-based device, with 32% of the theoretical capacity (based on PAQEDOT mass) available after five cycles.

**Conclusions**

This work resulted in the fabrication of 3D nanoarchitected, all organic, self-separating lithium ion EES devices. Devices were enabled by a co-continuously porous carbon anode, derived from the co-assembly of an ULMM BCP and resols. The next step was electropolymerizing a single-phase conductive cathode material in the form of the polymer PAQEDOT directly onto the porous carbon anode. Both sides were connected to leads, and through subsequent electrochemical processing in an electrolyte versus external lithium, an SEI was formed between the anode and cathode to act as a separator. While artificial SEI has been explored, this is, to the best of the author's knowledge, the first example of only a SEI being used as a nanoscale separator. After electrochemical processing to form the SEI and lithiating the cell electrodes, the EES device was pulled out of the electrolyte and cycled in the solid state. After 3 cycles with intermittent charges to 4.2 V, these devices output impressive discharge capacity despite little optimization. The process was also demonstrated to be generalizable to other co-continuous carbon materials and form factors, potentially opening applications with industrially scalable materials such as polymer membranes or carbon foams. Furthermore, these proof-of-principle devices can serve as a launching point for several research directions, including broadening the scope of 3D EES carbon anode morphologies and cathode chemistries, introducing the field of self-separating EES devices, and, more generally, highlighting the potential of all-organic material-enabled functional hybrid electronic devices.

    Despite these advances, significant challenges remain. The rapid capacity fade and limited charge rate, despite nm-scale critical dimensions, exhibited by our 3D battery devices underscore our incomplete understanding of the underlying electrochemical and transport mechanisms driving the performance of these devices. To that end, it is instructive to look at earlier reports on conductive redox polymer cathodes, using PEDOT as a backbone and quinones as pendant groups, showing appreciable cyclability in a half-cell (pouch cell form factor) against lithium metal, with a charge / discharge plateau around 3 V and 67% capacity retention after 200

cycles.[49] While this past research is not a perfect analog to the device detailed in our work, it does highlight the potential of our cathode material used and offers possible future directions of research to improve the cyclability of the devices described here. Wang et al. employed conducting redox polymer materials, where the quinone pendant groups were substituted with electron-withdrawing moieties; these substitutions both raised the formal potential of the polymers and onset potential for polymer conductance.[49] Their results indicated a strong association between polymer backbone conductance and pendant redox chemistry, and that a mismatch between conduction offset potential of the backbone and the redox potential for the second quinone reduction leads to trapped semiquinones and therefore incomplete conversion during reduction.[49] A similar issue could contribute to our device's low cyclability, with incomplete conversions leading to diminishing returns on each cycle. Exploring alternative pendant groups would be a future direction to improving cyclability, and conductance, since large pendant groups can potentially impede polymer backbone planarity and cause a higher doping level and onset potential for conductance.[50] Should a combination of pendant and conducting polymer backbone, as well as electrolyte, enable sufficient difference between the reduction potential of the quinone and the de-doping of the polymer backbone chain, then higher rate capabilities may also be unlocked due to bypassing sluggishness that may occur due to loss of conductance or incomplete quinone conversions. Due to the complex and integrated nature of the 3D EES device components, further optimizations in addition to cathode chemistry, such as electrolyte and SEI engineering, carbon structure of the anode (*i.e.* hard carbon versus graphitic carbon), and packaging of the device, may assuage current limitations on cyclability and power density.

**Acknowledgements**

W. R. T. T. gratefully acknowledges the Department of Energy for financial support through Grant No. DE-SC0010560. The authors would like to thank Prof. Héctor Abruña at Cornell for fruitful discussions of our work. This work made use of Cornell Center for Materials Research facilities. This work also used Cornell University NMR facilities, which are supported, in part, by the NSF through MRI award CHE-1531632.

Supporting Information

# All-organic self-separating three-dimensionally nanoarchitected electrochemical energy storage devices

William R. T. Tait[1,2], Sriram Murali[1], Chao-Hua Hsu[1], Jantakan Nedsaengtip[2], Christina Lee[1], R. Paxton Thedford[2], Jörg G. Werner[3,4], Vibha Kalra[2], Ulrich B. Wiesner[1,5,6,]*

1. Department of Materials Science and Engineering, Cornell University, Ithaca, New York, United States
2. Robert Frederick Smith School of Chemical and Biomolecular Engineering, Cornell University, Ithaca, New York, United States
3. Department of Mechanical Engineering, Boston University, Boston, USA
4. Division of Materials Science and Engineering, Boston University, Boston, USA
5. Department of Design Tech, Cornell University; Ithaca, New York, United States
6. Kavli Institute at Cornell for Nanoscale Science, Cornell University, Ithaca, New York, United States
* Corresponding author

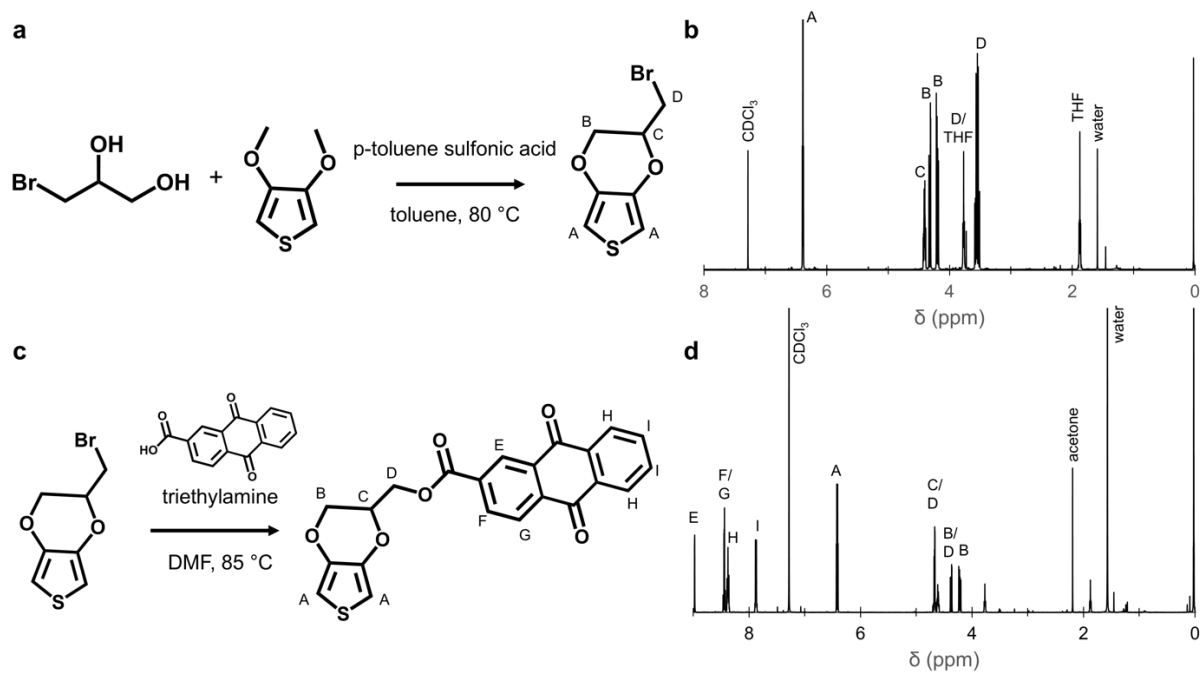

*Figure S1.* Reaction and proton NMR characterization of (a, b) Br-EDOT and (c, d) AQEDOT.

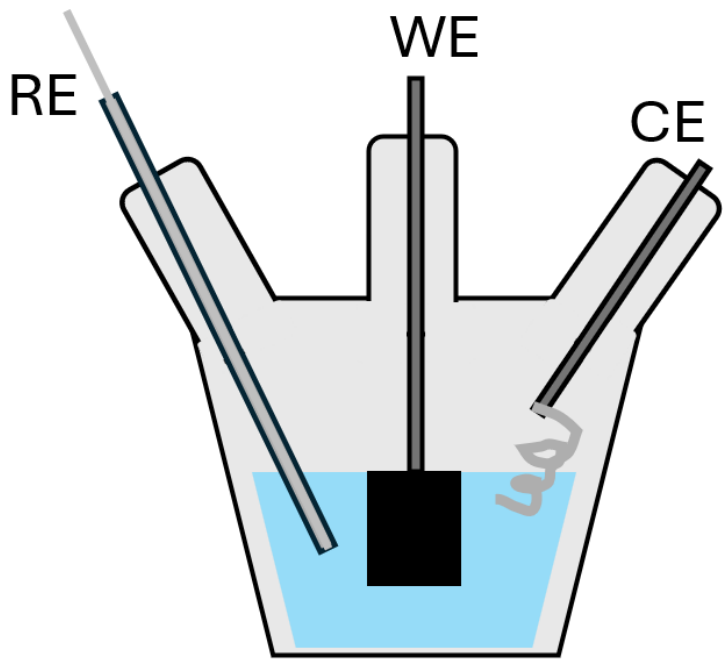

*Figure S2.* Schematic of custom glass electrochemical reactor set-up showing configuration for PEDOT current collector electropolymerization. RE a is Ag/Ag$^+$ reference electrode, WE is a carbon monolith working electrode, and CE is a platinum mesh counter electrode. Blue liquid is polymerizing electrolyte, 0.1 M LiClO$_4$ and 0.02 M EDOT in MeCN.

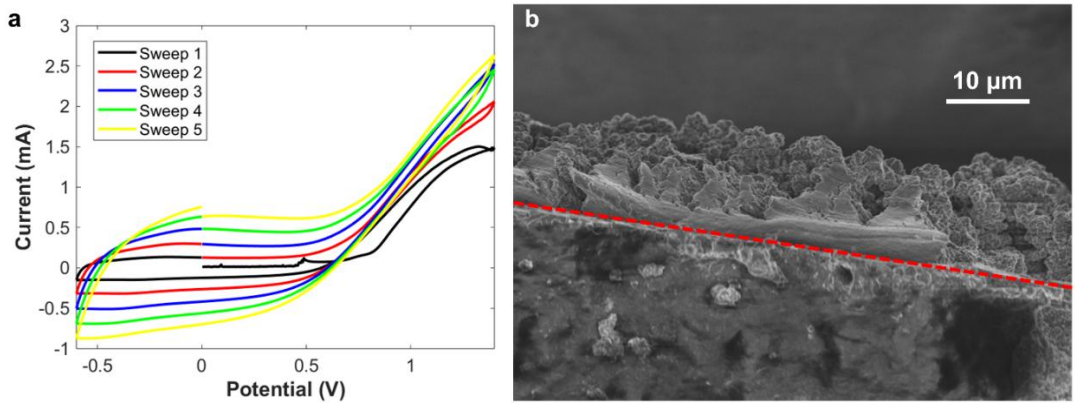

*Figure S3*. Deposition of PEDOT current collector. (a) CV of PEDOT surface current collector electropolymerization and (b) SEM image of PEDOT current collector on device surface, with edge between carbon and PEDOT delineated with dashed red line.

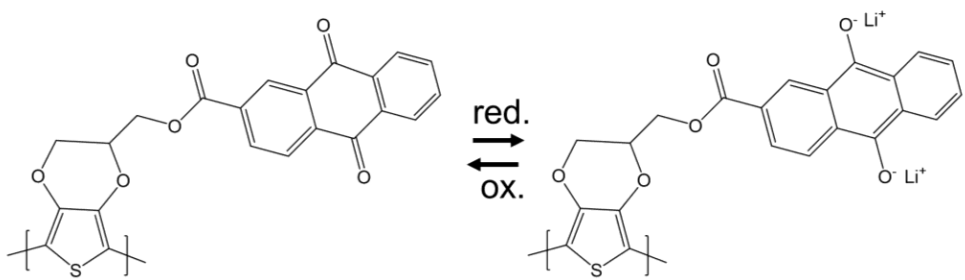

*Figure S4*. Oxidation and reduction of PAQEDOT anthraquinone pendant.

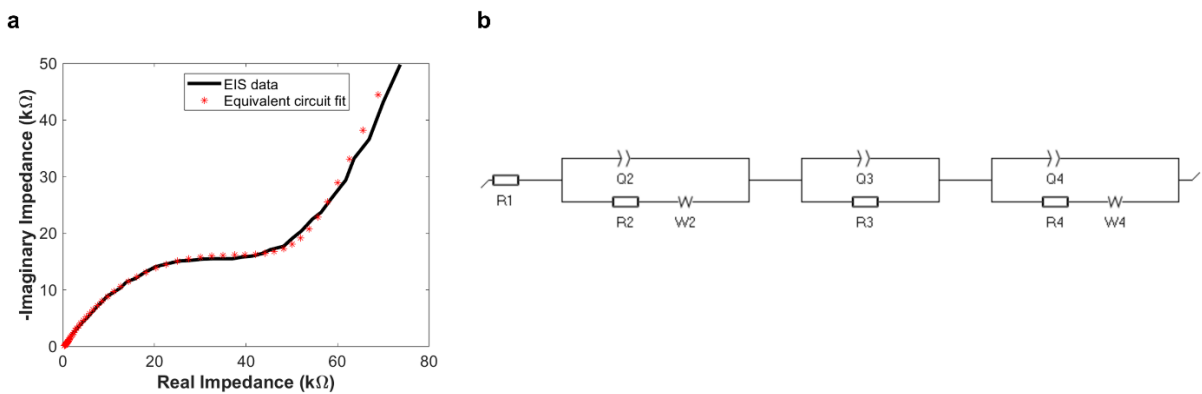

*Figure S5.* EIS data and equivalent circuit. (a) Nyquist plot of device from Figure 6. (b) Proposed equivalent circuit based on Nyquist plot fitting.

**Iteration of 3D battery design optimization**

Since more polymer per mass carbon seemed to potentially lead to improved performance, the BCP-based carbon was revisited and the deposition method of PAQEDOT was modified. In an effort to increase the amount of polymer deposited within the pore network, and potentially improve battery performance, a pulsed deposition method was explored, where 2.5 V vs AgQRE was pulsed for 1 second to polymerize the AQEDOT, the set-up was allowed to rest for 5 seconds to allow diffusion of monomer back into the pores where concentration may have depleted during oxidation, and, finally, a 5 second pulse at the reducing potential of -2.5 V was applied to draw in and reduce any leftover oxidized and non-polymerized monomer (Figure S7). The deposition voltage of 2.5 V was informed by the oxidation current of a CV deposition of PAQEDOT not beginning to increase until > 2 V, while the expected voltage is typically > 1 V vs Ag/Ag$^+$. The voltage shift, relative to an Ag/Ag$^+$ reference electrode, is anticipated to be from the use of an Ag wire as both pseudoreference electrode and counter electrode. This deposition method resulted in a BCP-based carbon backfilled with PAQEDOT with a 2:1 carbon to PAQEDOT ratio (vs. 3:1 from the original method).

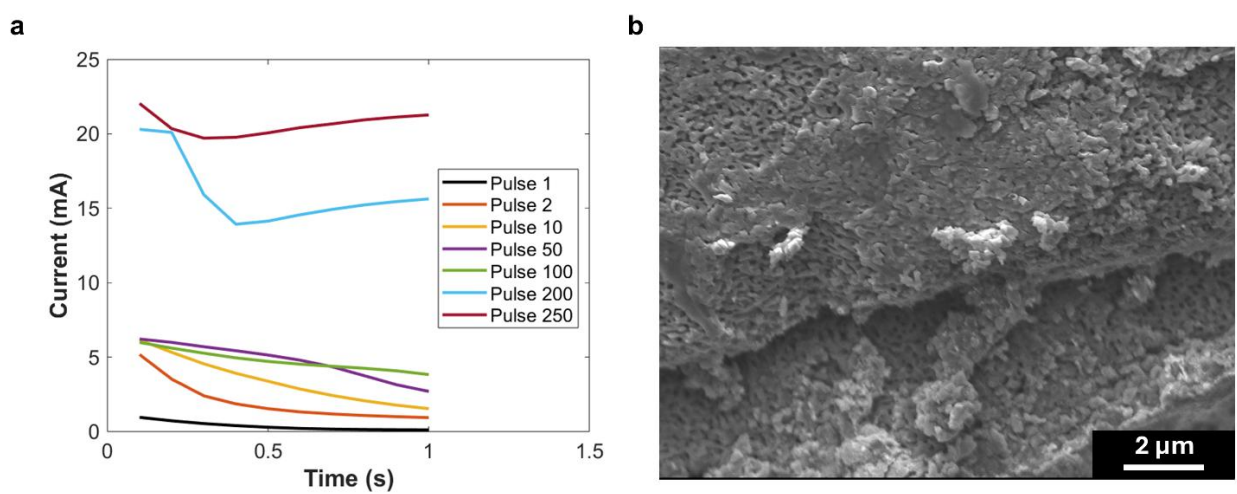

*Figure S6.* *Pulsed deposition of PAQEDOT. (a) Selected pulses traces of time versus current,*

*showing increasing current as conductive polymer is deposited. (b) SEM of cross-sectional fracture face surface of carbon post-deposition, showing pore space filled with polymer.*

Another condition changed for this next device was the addition of fluoroethylene carbonate (FEC) to the lithiating electrolyte, making the composition 1 M LiTFSI, 2 wt% LiNO$_3$, 2 wt% VC, and 2 wt% FEC. FEC was chosen as an additional additive for its potential to enrich the SEI with inorganic fluorinated species.[1] Other than the PAQEDOT deposition method and electrolyte composition, the same electrochemical processing procedure was used to convert the hybrid material into a functional EES device (Figure S7).

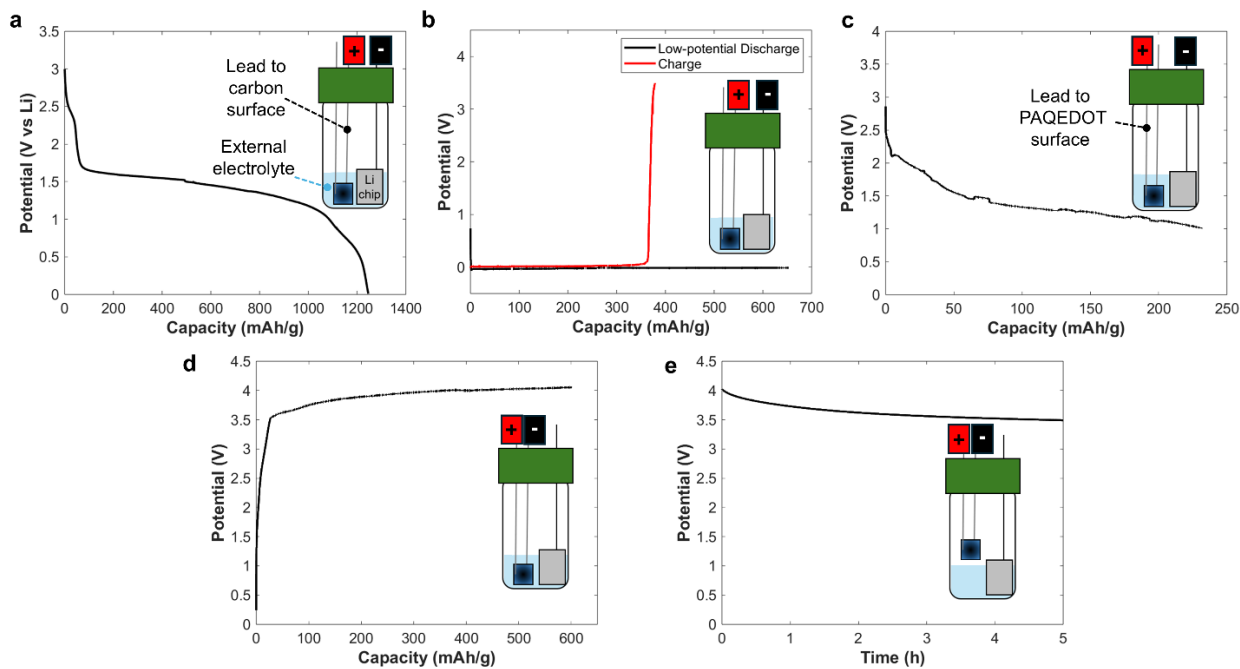

*Figure S7. Electrochemical processing of BCP carbon-based 3D EES device with FEC included in lithiating electrolyte. (a) SEI generation. (b) Low potential subsequent discharge for further SEI generation followed by charge to strip plated lithium. (c) PAQEDOT lithiation. (d) Charging of device in electrolyte. (e) Device holding open circuit potential.*

Furthermore, the higher-than-theoretical discharge capacity of the third discharge for the membrane-templated carbon-based EES device, combined with the small dip in the profile of the third discharge around 1.8 V, led us to select 1.8 V as the cut off discharge voltage for this subsequent device cycling test. The results for the 3D EES full cell device in Figure 6e (see Main Paper) showed very large discharge capacities, with the third cycle discharge capacity being about 267 mAh/g, well beyond that of the theoretical capacity of PAQEDOT while just considering the charge compensation of the reduced carbonyl oxygens. After the third cycle, once again, significant capacity fade occurred. In addition, a second device was made using the pulsed deposition method and an FEC-rich electrolyte, which confirmed these results (Figure S8). The increase in capacity could be due to contributions from PEDOT, which itself is pseudocapacitive and has demonstrated appreciable capacity.[2] If both the PEDOT backbone of PAQEDOT and the PEDOT surface current collector contributed to the capacity, this may explain the large gravimetric specific capacities based solely on the weight of the PAQEDOT in the device.

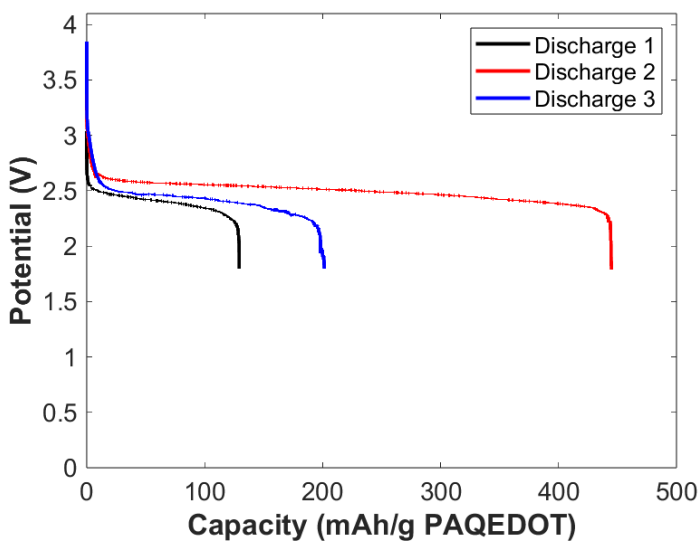

*Figure S8.* *Cycling data for additional BCP carbon-based 3D EES full cell device to repeat results.*

While the charge portion of each cycle was set to a cut-off voltage of 4.2 V, like in the device from Figure 6 (see Main paper), the early charges had long plateaus at around 4.05 V followed by unstable voltages rather than reaching 4.2 V (Figure S9). They were cut off early since the large charge capacities and unstable voltage under constant current were indicative of some potential side reaction. Later cycles reached 4.2 V without needing to be manually cut off.

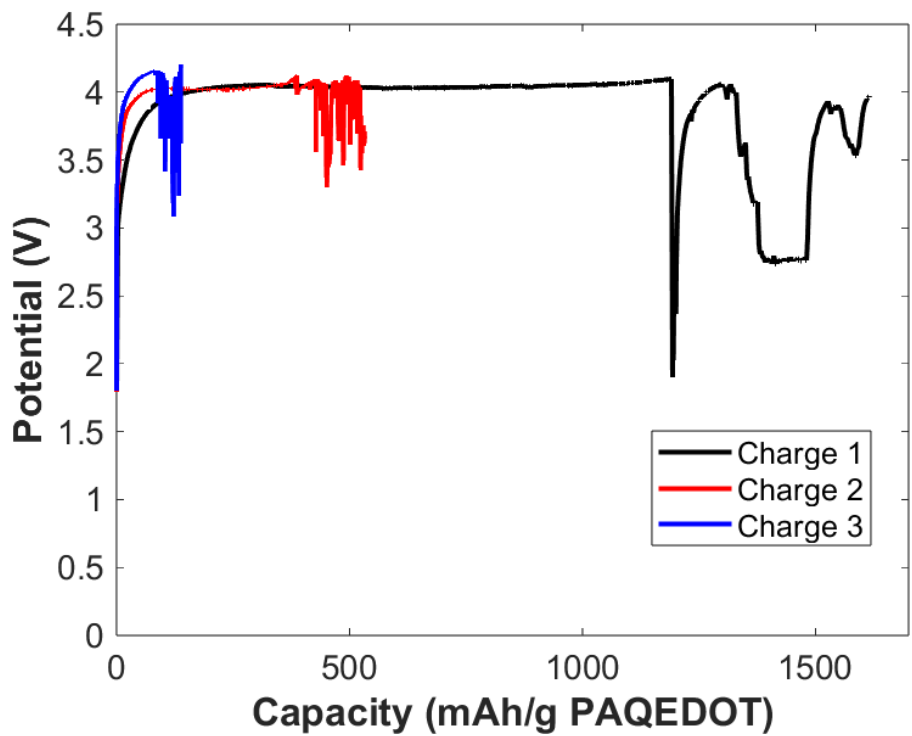

*Figure S9. Charge profiles from constant current cycling of device from Figure 6e, Main Paper.*

To test the potential contribution of PEDOT, a device was fabricated using PEDOT instead of PAQEDOT for the initial deposition, as well as a PEDOT surface current collector. When the all-PEDOT cathode EES device was cycled (Figure S10a), early cycles had appreciable capacity while the capacity fell quickly beyond cycle 3. Furthermore, rather than flat faradaic plateaus (Figure 6e in the Main paper, Figure S8), the discharge profiles were sloped, indicative of a pseudocapacitive mechanism. Cyclic voltammetry of the full device also pointed towards capacitive behavior, with the CV profile having a featureless rectangular shape that increased proportionally in area with scan rate (Figure S10b, Figure S10c). Another possible explanation was some degradation of the cathode material or SEI resulting in the extra capacity.

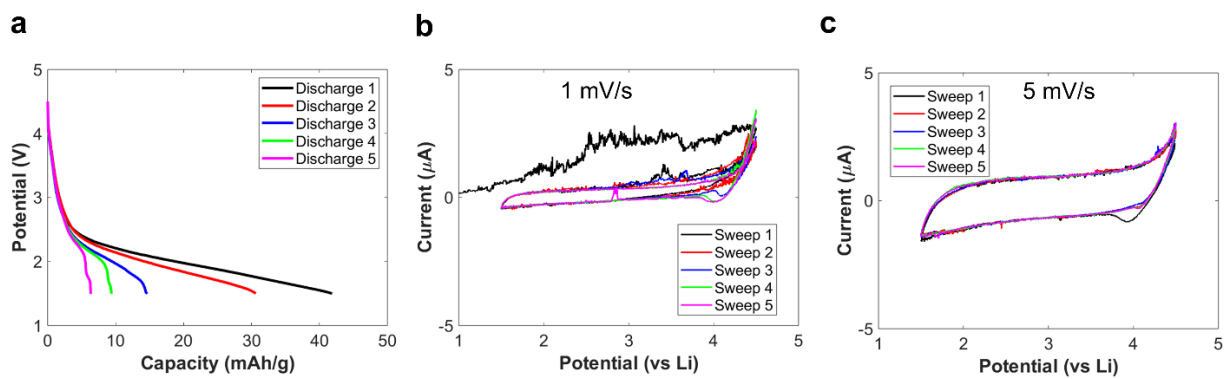

*Figure S10.* *Cycling and CV data for pure PEDOT cathode 3D EES device. (a) Discharge capacity of first five cycles. (b, c) CVs of 3D EES device at 1 mV/s and 5 mV/s, respectively.*

In an attempt to better understand the operation window of the device, a CV of a thin film of PAQEDOT on a glassy carbon electrode was taken in 0.1 M LiClO$_4$ in acetonitrile (Figure S11) at a rate of 0.5 mV/s. Over three sweeps, this CV showed reduction peaks at approximately -0.94 V and in a range of -1.86V to -2 V. These peaks are expected to correspond to the reduction of the anthraquinone (AQ) pendant of the PAQEDOT and the de-doping of the PEDOT backbone, respectively.

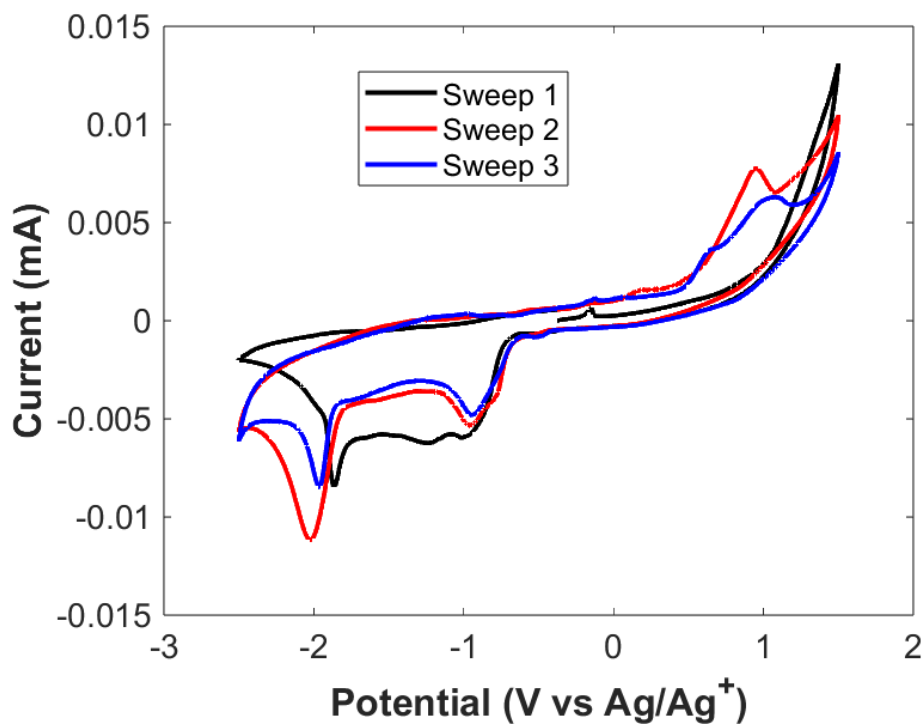

*Figure S11.* CV of thin film of PAQEDOT on glassy carbon electrode in 0.1 M LiClO$_4$ in acetonitrile.

There is also decreasing area under this peak at ~0.94 V with each sweep, indicating some irreversibility to the process which could impact PAQEDOT conductivity. The reduction peak around -1.9 V has a less consistent position, ranging from -1.86 to -2 V vs. Ag/Ag$^+$. The inconsistency in peak position and intensity could have to do with overpotential originating from the incomplete doping or de-doping of the PEDOT backbone, which could, in turn, impact the charge transfer for the redox component of the AQ pendant. The oxidation peak at ~0.95 V versus Ag/Ag$^+$ having inconsistent peak area and location also supports this notion, which may explain the quick capacity fade of these devices. This indicates that with further optimization, it may be possible for there to be less capacity fade and more cyclability in these devices. While the cyclability was low, these proof-of-principle full cell EES devices still reached non-trivial discharge capacities in the solid state, showing the promise of taking on energy storage and devices from an architectural perspective. The nonconventional architecture of the design as well as the all-organic materials choice allowed for a first-of-its-kind manufacturing strategy with in situ separator generation by the SEI.

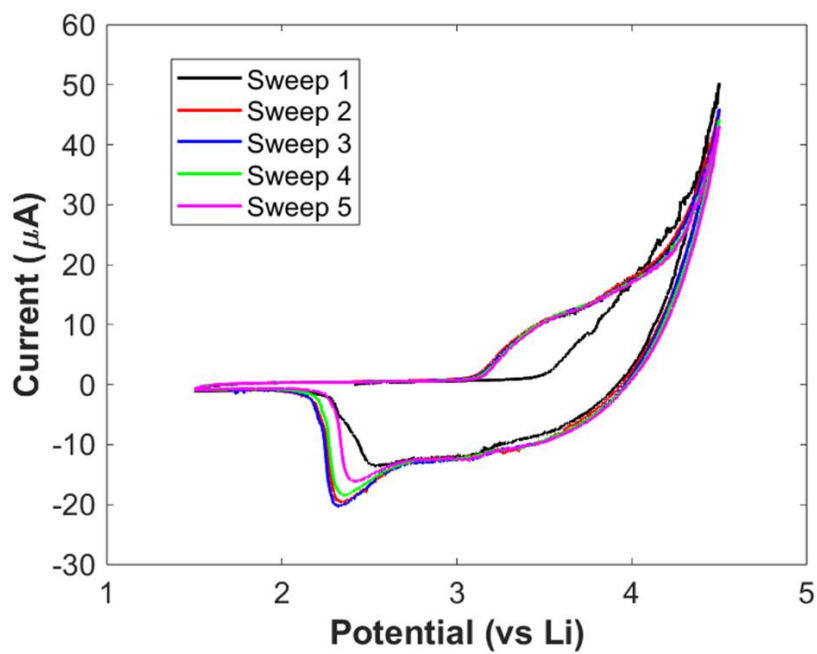

*Figure S12.* CV of device from Figure 7 at 1 mV/s.

**SI References**